\documentstyle[12pt,aasms4]{article}
\slugcomment{AJ manuscript, \ \ revised version for submission, 03 August 95}
\lefthead{Zacharias et al.}
\righthead{A Radio--Optical Reference Frame VIII}
\begin{document}

\title{A Radio--Optical Reference Frame \\
       VIII. CCD observations from KPNO and CTIO:\\
       internal calibration and first results}
\author{N. Zacharias\altaffilmark{1},
	C. de Vegt\altaffilmark{2},
	L. Winter\altaffilmark{2},
	K.J. Johnston}
\affil{U.S. Naval Observatory, 3450 Mass. Ave. N.W., Washington D.C. 20392}

\altaffiltext{1}{with Universities Space Research Association (USRA),
  Division of Astronomy and Space Physics, Washington D.C.}
\altaffiltext{2}{Hamburger Sternwarte, University of Hamburg, F.R. of Germany}

\begin{abstract}
In this pilot investigation,
precise optical positions in the FK5 system are presented
for a set of 16 compact extragalactic radio sources, which will
be part of the future radio--optical reference frame.
The 0.9 m KPNO and CTIO telescopes equipped with 2K CCD's
have been used for this project.
The astrometric properties of these instruments are investigated in detail.
New techniques of using wide field CCD observations for astrometry
in general are developed.
An internal precision of 5 to 31 mas in position per single exposure is found,
depending on the brightness of the object.
The tie to the primary optical reference system is established by
photographic astrometry using dedicated astrographs on both hemispheres.
An accuracy of $\approx 30$ mas per source is estimated for the multi--step
reduction procedure when based on the future Hipparcos catalog,
while the FK5--based positions suffer from system errors of 100 to 200 mas
as compared to the radio positions.
This work provides a contribution to the international effort
to link the Hipparcos instrumental coordinate system to the
quasi--inertial VLBI radio reference frame.
Precise radio and optical astrometry of a large sample
of compact extragalactic sources will also contribute to
the astrophysics of these objects by comparing the
respective centers of emission at the optical and radio wavelengths.
\end{abstract}

\keywords{astrometry: reference frame, CCD observation and reduction techniques
	  --- Hipparcos: extragalactic reference frame link
	  --- QSO, BL Lac: optical positions}

\section{Introduction}

This paper is part of a series of papers describing the
construction and maintenance of a quasi--inertial reference frame
in both the radio and optical domain.

Extensive work on the radio reference frame has been accomplished
in the last 5 years and published in a series of papers I--VII
(see Johnston et al. \markcite{rmRORF} 1995 for further details).
A rigorous global new reduction of all then applicable Mark III VLBI
radio observations has been used to construct a radio reference frame
of 560 sources from first principles (Johnston et al. \markcite{rmRORF} 1995).
Based on these results a list of
defining and candidate sources has been provided to the IAU
Working Group on Reference Frames (IAU, \markcite{rmIAUWG} 1995).
While a dense radio frame with an accuracy level of 1
milliarcsecond (mas) for most of the source positions is now
in place, optical observations on a 30--50 mas level are available
for only a fraction of these sources.

Previous results already have shown the deficiencies of the
currently used optical FK5/J2000 reference frame, with deviations
from a uniform inertial reference system as large as $\approx 200$ mas
at the current epoch.
The Hipparcos astrometry satellite mission will soon provide
a new optical system on the 2 mas level
at the Hipparcos mean epoch ($\approx 1991.5$),
and 2 mas/year in proper motion error,
but this instrumental
system must be linked to the radio system to become quasi--inertial.
Most of the primary optical reference objects (stars)
are bright, both in the FK5 (3 to 6 mag)
and the Hipparcos (5 to 9 mag) catalogs, while the optical counterparts
of the extragalactic sources are optically faint (the majority in the
range 17 to 21 mag).

Due to the relatively low quantum efficiency of photographic astrometry,
which requires long exposure times on large telescopes in good seeing,
progress in the optical observations has been slow.
The use of CCD detectors has dramatically improved this situation,
because they allow smaller telescopes which have greater availability
and more objects can be observed due to much shorter exposure times.
Not until recently has the fieldsize of CCD's became large enough to
contain a sufficient number of reference stars for precise astrometry.

This paper outlines the reduction procedure in detail and gives
results for a representative subset of the observed sources.
It is a pilot investigation to assess the astrometric properties
of these telescopes and the capabilities of this technique.
The first successful attempt to use CCD's for this project has been made
earlier (de Vegt \markcite{rm1CCD} et al. 1987),
although it was severely limited by the lack
of reference stars in a tiny 2' by 3' field of view.

In Section 2 we discuss the telescopes, CCD's and observations
and in Section 3 the reference star data.
Section 4 describes the reduction procedure, while results are
presented in Section 5.
An accuracy estimate of the procedure and comparison with
other investigations is made in Section 6.

\section{OBSERVATIONS}

\subsection{The Telescopes}

The 0.9 m (36 in) Kitt Peak National Observatory (KPNO)
telescope is a Ritchey--Chr\'etien system with an additional
2--element field corrector about $250 \ mm$ before
the focal plane.
This gives a large ($\approx 1\deg$) flat field of view,
which also gives the offaxis guide scope good image quality.
Frequent focus measurements and the use of the correlation
between focus setting and telescope
temperature ensured optimal image quality for all
object frames.
The image quality at the KPNO instrument was found to
be uniformly good over the entire field of the CCD.

The 0.9 m (36 in) Cerro Tololo Inter--American Observatory
(CTIO) telescope is a cassegrain system without a
field corrector. At the edge of the CCD frame optical aberrations
are visible and it is difficult to achieve a well focused CCD frame with
round images over most of the chip area due to instabilities in
the mirror--supporting structure.
We also had occasional guiding problems.
On the positive side the longer focal length of the CTIO 0.9 m with
its better sampling, makes it more suitable for structure analysis of
the objects than the KPNO 0.9 m. Getting enough reference stars for the
astrometric link was found to be the bottleneck with the CTIO 0.9 m.

Properties of both telescopes and the CCD's used
are summarized in Table 1.

\subsection{The Data Acquisition System}

Both telescopes use the same type of Tektronix 2K CCD chip,
with square pixels of size $24 \ \mu m$ and a filling factor of 100\%,
with a different camera controller.
For good astrometric results, a large
S/N ratio is required, whereas an optimized
digitization at the background level is of minor importance.
In order to cover the large magnitude
range between the reference stars and the extragalactic sources,
a large gain of 8.2 was chosen for the KPNO instrument
to utilize the full dynamic range, including the full well
capacity of the chip.
For the CTIO camera, a gain of 3.3 was sufficient because of
the larger digitization range (16 bit) available with that camera controller.

The readout time for the KPNO instrument was well over
2 minutes. The new ARCON controller at CTIO
allowed a faster readout of the full frame in about 70 seconds
with two readout amplifiers.

All frames have been taken in a red spectral bandpass.
A Gunn r filter was used for the long exposure frames
to record the extragalactic objects.
For each object, additional short exposure frames
have been taken in order to get unsaturated images of the
reference stars (12 to 14 mag).
With the KPNO 0.9 m, most of these short exposure frames have been
taken with a narrow (FWHM = 12 nm) filter centered near $H\alpha$.
The better sampling of the CTIO instrument
allowed the use of the same Gunn r filter for the short exposure frames
because the flux of the secondary reference stars was spread out over
more pixels, thus avoiding saturation.

The IRAF software system was used for the data acquisition at both sites.

\subsection{Observation Procedure}

A summary of the observing runs is presented in Table 2.
Between April and October 1994 the dome seeing was improved
at the 0.9 m KPNO telescope.
In the first observing run at each site many calibration and test
frames were obtained in order to evaluate the
astrometric quality and possible systematic errors
of the instrumentation, as well as to determine the
best observing strategy.

The second runs at each telescope were pure production runs,
taking at least 2 long and 2 short exposure frames per field.
Depending on the brightness of the sources,
an exposure time of 200 to 900 seconds was used for the deep
frames and 40 to 120 seconds for the others.
At the KPNO telescope
a typical deep frame covers an astrometrically useable range of
$14 \ to \ 20^{m}$, while the short exposures cover $10.5 \ to \ 16.5^{m}$.
At CTIO the corresponding ranges are
$13 \ to \ 20^{m}$ and $11 \ to \ 18^{m}$.
All frames were taken within 1 hour of the meridian and
at least $30\deg$ away from the Moon.

In fields of high galactic latitude the density of the
secondary reference stars are usually not sufficient for
a good astrometric reduction.
In those cases a mosaic of short exposures (2 x 2 frames),
centered on the object and shifted by 500" in $x$ and $y$
were obtained at KPNO.
The overlap of 41\% in area
allows a rigid tie to the central frames and
the area covered for potential reference stars
was increased by a factor of 2.5.
Because of the much smaller useable field size of the CTIO
0.9 m, a similar 2 x 2 mosaic with shifts of 240" in x
and y were obtained for all fields at that telescope.
This is an overlap of 50\% by area.

A few additional calibration fields were taken
with at least a 3 x 3 set of frames of 2--3 minutes
exposure time.
On some nights, additional sets of short exposure
(10, 20, 40 seconds) frames
were obtained in order to investigate the
limits set by atmospheric turbulence on astrometric accuracy.
Results will be published elsewhere (Zacharias \markcite{rmZatm} 1996).

It was necessary to have two observers present,
in order to obtain online quality control.
On each deep frame, the object was identified
and radial profile and contour maps were generated.
A few objects were found to be optical doubles on the
2 to 7 arcsecond level, most likely due to foreground
stars.
Depending on the seeing, these objects were skipped or
the exposure time was adjusted, if required.
All frames were checked for focus and overall image
quality.
The short exposures in addition were checked
for saturation of the reference star images and
the exposure time was adjusted accordingly.

\section{Reference Stars}

The currently used primary reference system is the
IRS (International Reference Stars),
(Corbin \& Urban \markcite{rmIRSp} 1990, Corbin \& Warren \markcite{rmIRSc}
1991).
The IRS gives positions and proper motions in the FK5/J2000 system
for approximately 36,000 stars in the magnitude range of
$V \approx 6 \ to \ 9$, nearly uniformly distributed on the sky.

Nearly all radio source fields from our candidate list
(Johnston et al. \markcite{rmRORF} 1995)
already were observed with modern high precision astrographs.
The northern hemisphere plates were taken with the 23 cm Hamburg
Zone Astrograph (ZA) (de Vegt \markcite{rmCDS} 1978) and the southern
hemisphere
plates were obtained with the yellow lens (BY) of the U.S. Naval
Observatory 8 in Twin Astrograph (Routly \markcite{rmTA} 1983)
from the Black Birch Astrometric Observatory (BBAO) in New Zealand.
The field sizes of these instruments are $ 6\deg \times 6\deg$ (ZA)
and $ 5\deg \times 5\deg$ (BY) respectively.
Both instruments use $6^{m}$ objective gratings in order to obtain
diffraction images of bright stars for high precision position
measurements.
The range of magnitudes covered is $V \approx 5 \ to \ 14$
in 15--minute exposures on microflat 103aG emulsion.
Both instruments have 2 meter focal length, corresponding to
a plate scale of about $100"/mm$.

All IRS and Hipparcos Input Catalog (HIC) stars
(Turon et al. \markcite{rmINCA} 1992)
have been measured on these plates together with all faint
stars to the plate limit in an area of $1\deg \times 1\deg$
centered on the radio source position.
These stars in the magnitude range of $V \approx 11 \ to \ 14$
serve as secondary reference stars for the reduction of
the CCD frames.

In addition, the 0.5 m Lick Carnegie Astrograph (LA) was used
to provide a denser net of secondary reference stars in selected fields.
The LA has a field size of $ \approx \ 3\deg \times 3\deg$
on $240 \ mm \ \times \ 240 \ mm$ plates
but a limiting magnitude of $V \approx 15.5$ in 30--minute
exposures with a plate scale of $55"/mm$.
For many fields on the southern hemisphere, plates from
the ESO Schmidt telescope are available, which often show
measurable images of the radio sources.
All Schmidt plates provide at least a set of tertiary
reference stars in the magnitude range of $R \approx 14 \ to \ 18$.
Measuring of all those plates is in progress
at Hamburg Observatory with the HAM--I machine
(Winter et al. \markcite{rmW1} 1992, Winter \markcite{rmW2} 1994,
Zacharias et al. \markcite{rmZ161} 1994).
An accuracy of $\approx 0.8 \ \mu m$ per coordinate is obtained for
the measurement of a single image on a good astrograph plate.

For the present pilot study only a subset of CCD observations were
used to obtain positions of the optical counterparts of the
extragalactic sources.
A publication for all fields is in progress and will
be based on the Hipparcos catalog.

Ultimately all CCD frames of the 0.9 m telescopes depend
entirely on the astrograph observations,
which provide the high precision secondary reference star positions.
The Guide Star Catalog (GSC) is not precise enough
for this project and results from CCD transit circle
instruments are premature at the moment.
The future Tycho catalog alone will not be dense enough
to provide a good reduction of the current CCD observations.

\section{Reduction Procedure}

\subsection{Calibration of Raw CCD Frames}

The standard IRAF software package
(version 2.10, NOAO, Univ. of Arizona, Tucson)
was used for the initial reduction steps of the raw CCD data.
About 20 bias frames were combined for a master Zero for each night.
Most of the KPNO frames were calibrated with flatfields derived from
a large number of object frames from a group of nights in which the
dust grain pattern on the filter and dewar window remained constant.
CCD frames taken with the narrow filter at KPNO and all CTIO frames
were calibrated with twilight flats.

\subsection{Pre--processing Statistics}

Statistical information such as
noise characteristics of the frame and full width half maximum (FWHM)
values of the image profiles were obtained from all calibrated images
using standard IRAF commands.
An approximate pixel position and count rate of the extragalactic objects
were obtained from radial profile fits, as well as relative position offsets
for the short exposure frames.
This information is read by our reduction programs for
an automatic data handling of multiple frames and fields.

\subsection{Determination of $x,y$ Coordinates}

In order to obtain centroid positions of star images ($x,y$ coordinates)
two different software packages were investigated for comparison.
First, the IRAF/DAOPHOT routines provided three different sets of
$x,y$ coordinates:
\ a) a center of mass position,
\ b) a 1--D gaussian fit to the $x,y$ marginal distribution of the pixel data,
and \ c) a point spread function (PSF) fit after subtraction of a gaussian.
Only options a) and b), provided by the routine {\em phot}, include
an error estimate on the derived $x,y$.
Option c) is obtained with {\em allstar} (Stetson \markcite{rmDAOPHOT} 1987)
which can handle crowded fields
by a simultaneous fit of profiles to a group of stars and gives
superior photometric results, but is not designed for astrometry.
Second, {\em SAAC} (Software for Analysing Astrometric CCD's) was used.
SAAC was developed at Hamburg Observatory
(Winter \markcite{rmW2} 1994) with various modifications and adaptions
performed by
N.Zacharias at USNO and
performs 2--D fits on the star profiles with a choice of various models.

Slightly saturated images of bright stars usually are of good astrometric
quality
as long as there is no bleeding into adjacent pixels.
These images were kept, with caution, in the following reduction steps.
Both program packages exclude significantly elongated images, mostly galaxies
and various defects due to restrictions imposed on various image selection
parameters.

Positions of radio sources presented in this paper
are based exclusively on the circular gaussian 2--D fit,
which proved to be the best option for astrometry.
For optical doubles the same fit model was used
either with the cut--out or with a fit--up procedure (Schramm
\markcite{rmSdiss} 1988) or both.
In the cut--out option all
pixels containing the companion image are excluded from the profile fit
observation equations (interactively selected).
In the fit--up procedure the background level is raised to a higher
value up to the maximum count rate from the companion within the
used window.
Bright companion star images were first fitted by these methods exclucing
the image of the etragalactic source.
The thus obtained image profile was subtracted from the original
pixel data and finally the position of the extragalactic source determined.
The {\em allstar}(DAOPHOT) procedure was used for comparison in some cases.

\subsection{Comparison of $x,y$ Data}

Unique star numbers were assigned to all images of all stars of
each field using our multi--step match program, based on position only.
Coordinates were corrected for approximate third--order optical
distortion, derived from a pilot investigation of a subset of the frames.
Then all matches within a large search radius
are recorded and the distribution of the coordinate differences
of the matches in x and y are analyzed for a peak.
Only the most likely identified matched
images are used for a linear transformation using an unweighted least
squares adjustment.
Finally the transformed coordinates are compared to the
reference coordinates and a position match with small tolerance is performed.
Optionally, all images identified as multiple on a few arcseconds level are
rejected except for manual 'fine tuning' for some radio source images.

A transformation program was developed for comparing
$x,y$ data of all overlapping CCD frames of the same field.
Various mapping models are available and
the residuals can be analyzed and displayed with already
existing software from the photographic plate reduction package.
The transformation program was used for comparing
$x,y$ coordinates of the same CCD frame as obtained by
different pixel fit algorithms, as well as to estimate
field distortions and positional accuracy.
Optionally, the obtained transformation parameters were applied
to {\em combine} the $x,y$ data of all frames of each field
into a superframe, from which spherical coordinates could be derived.

\subsection{Combining $x,y$ Data}

One option to derive positions from the combination of all CCD frames
of a field is the traditional way of adjusting
each individual CCD frame's $x,y$ data to spherical coordinates
($\alpha, \delta$) by means of reference stars and then combining these
to obtain (weighted) mean positions.
This option is not likely to give best results here.
The number of available reference stars is usually small.
Also, the deep exposure frames have overexposed images
of the reference stars and an iterative process involving tertiary
reference stars' $\alpha, \delta$ then is required, starting from
the short exposure frames.

Combining the $x,y$ data first is feasible
without loss in accuracy due to the small field of view,
the simple transformation geometry required here,
and the large overlap in area of $\approx 40 \ to \ 100 \%$ for all
our CCD frames of a field.
Our software allows for a rigorous correction for refraction,
but this effect was found to be negligible for our data.
A significant third--order optical distortion term was
removed before the weighted least squares adjustment with
a linear transformation model for overlapping frames.
Weights were obtained from the precision of the image profile
fits, which strongly depend on the magnitude of the stars.
In addition a constant variance per frame was added to account for
atmospheric effects, scaled by the inverse exposure time
of the CCD frames.
This approach allows using simultaneously more reference stars
in a larger field of view in cases when mosaic CCD frames are
available, and at the same time combines all measurable images
of long and short exposure frames into a common system.

\subsection{Adjustment to Reference Star Positions}

Our standard plate reduction software package, as part of HBAPP
(Hamburg Block Adjustment Program Package, Zacharias \markcite{rmZdiss} 1987),
was used for the unweighted least--squares adjustment of the $x,y$ data
to the reference star positions.
The precision of the secondary reference stars contributes the largest
part of the errors in this adjustment. The differences in the precision
of the CCD $x,y$ data, e.g. due to the dependence on magnitude or
the atmospheric effects as a function of exposure time, are not
relevant here.
Experiments with weights obtained from the astrograph reductions
did not reveal any significant improvement.

Several pilot investigations were run to determine the most realistic
mapping function of the telescopes (plate model).
A linear plate model was finally adopted for routine processing with
third--order optical distortion corrected prior to the adjustment.

\subsection{External Comparisons}

Field, as well as magnitude--dependent, systematic
errors can be investigated externally by comparing
$\alpha, \delta$ coordinates obtained from our CCD frames
with positions obtained from a different telescope.
For some fields ESO (European Southern Observatory) Schmidt
plates, Lick Astrograph plates or prime focus plates from other
telescopes are available.

Due to the high precision of our CCD data a comparison with
Schmidt plates only reveals systematic errors in those plates.
Positions obtained from prime focus plates depend on the same
secondary reference stars of that field already used for the CCD
reductions.
Thus the Lick Astrograph offers the most promising external
comparison. Plate measuring is in progess and results
will be published in an upcoming paper.
Here only a comparison between different KPNO and CTIO runs
will be made.

A comparison of the optical positions with the quasi--error--free
radio position can also reveal systematic errors.
But our aim is to calibrate the optical data independently of that
comparison in order to draw astrophysically significant conclusions.

\section{Results}

Because no photometric observations were obtained, all magnitudes
in this paper are instrumental with the zeropoint adjusted to the mean
photographic V magnitudes of the secondary reference stars.

\subsection{Internal $x,y$ Precision}

In this section we will compare the astrometric performance of
various image profile fit models
used to obtain $x,y$ coordinates from the pixel data of the CCD frames.

Initial tests with the 1--dimensional center of mass algorithm
showed significantly larger errors in positions as compared to other
algorithms and thus was not further investigated.
We compared the 1--dimensional gaussian fit (1DG)
and the {\em allstar} point spread function fit (ALS) from the DAOPHOT/IRAF
package, as well as the 2--dimensional circular gaussian fit (2DG)
from our astrometric software package (SAAC).

All comparisons were made only with stars appearing
on all lists of $x,y$ coordinates obtained for the different profile fit models
per frame, thus a unique limiting magnitude was used.
All $x,y$ data were corrected for third--order optical distortion
by applying the same value for the distortion coefficient (D3) and the
location of the optical axis on the frame ($x_{0}, y_{0}$) to all frames prior
to the transformation with a linear model.

For routine reduction of the CCD frames in order to derive positions of
extragalactic sources, the 2--dimensional circular gaussian (2DG) was adopted
as our standard fit model because of its superior performance with
respect to random, as well as systematic, errors as will be explained in the
following two subsections.

\subsubsection{KPNO 0.9 m Telescope}

Table 3 shows some results for the KPNO field of the source 1656+053.
The CCD frames with internal numbers 68 and 74 are long exposures (300 sec),
while all others are short exposures (40 sec).
Frames 68,74 and 73 are centered on the QSO radio source,
while all others are part of a mosaic with offsets of 500" (735 pixel)
in each coordinate.
The seeing for all frames was $FWHM \approx 2.5$ pixel.

Most of the $x,y$ data obtained from the 1DG fit
show a significant magnitude--dependent error
in both coordinates for the faint stars as compared to the $x,y$ positions
from the same CCD frame obtained by the other fit models.
An example is shown in Figure 1 a) for the x coordinate of frame 73
in the comparison of fit model 2DG - 1DG.
The other coordinate, as well as the data from other frames, looks similar.

The average position difference, as obtained by different profile
fit models for the same pixel data, is on the order of 0.01 to 0.03
pixel, which is about 7 to 20 mas for the KPNO telescope.
This is clearly a function of magnitude, increasing to the faint end.
Figure 2 a) shows an example of the fit precision, $\sigma_{fit}$, as a
function
of magnitude for the 2DG fit of the pixel data obtained at KPNO from a
300--second exposure CCD frame of the field 0906+015.
The saturation limit for this frame is at $\approx 13.0^{m}$.
A best positional precision of 0.01 pixel is found for star images
in the magnitude range from the saturation limit to 3 magnitudes below.
For faint stars the fit precision decreases sharply.
Outliers have been visually inspected on the CCD frame and only
galaxies and close double stars have been found to cause
significantly larger fit errors than the mean for that magnitude,
providing a good method for detecting non--stellar images.

Positions from multiple CCD frames of the same exposure time and the
same field center were compared next.
The standard error $\sigma_{xy}$ of such a frame--to--frame position
transformation includes, besides the fit error $\sigma_{fit}$,
also errors introduced by the atmospheric turbulence $\sigma_{atm}$.
Figure 3 a) shows an example of $\sigma_{x}$ vs. magnitude
for 2 frames from the field 0906+015 with a 300--second exposure time each.
The limiting precision of $\approx 5 $ mas
can entirely be accounted for by the turbulence in the atmosphere.
According to Lindegren \markcite{rmLin} (1980)
we have for our case $\sigma_{atm} \approx 10 $ mas.
Compared to the previous figure either $\sigma_{fit}$ or $\sigma_{atm}$
or both are overestimated.
An explanation for overestimating $\sigma_{fit}$ for bright stars is
the difference of the observed image profile as compared to the assumed model.
On the other hand, significant nightly variations of $\sigma_{atm}$
are also well known.

Positions obtained for the brighter stars have been found to be less dependent
on the profile fit algorithm used than those obtained for fainter stars.
There was no systematic radial difference vs. radius found in
any fit model comparisons of the same KPNO CCD frame.

The transformation of $x,y$ coordinates of the frames with short exposure times
show a larger sigma than those of long exposure times for all fit models
(Table 3) because of the noise added by the atmosphere.
In comparing the performance of the different fit models,
the 2DG shows the smallest random errors for the KPNO frames.

\subsubsection{CTIO 0.9 m Telescope}

Table 4 shows some of the results for the CTIO telescope for the field
0646--306.
Frames 53 and 54 are long exposures (600 sec, 300 sec), while all others are
short exposures (40 sec). Frames 53, 54 and 57 are centered on the QSO,
while all others are part of a mosaic with offsets of
240" = 600 pixel in both coordinates.
Frame 53 has the poorest image quality as compared to the other frames.

Contrary to the results obtained with the KPNO telescope, here the simple
1DG fit is in good agreement with the 2DG fit.
An example is shown in Figure 1 b) for the difference in x--coordinates vs.
magnitude for frame 57.
No magnitude--dependent systematic errors were found in the test field.
The CTIO telescope has better sampling
and with FWHM $\approx 3.8 \ px$ in this test field there is a sufficient large
number of pixels for both algorithms to determine consistent positions
over a dynamic range of almost 8 magnitudes.

With the CTIO data the DAOPHOT {\em allstar} algorithm
show small magnitude--dependent systematic differences ($\approx 10 $ mas/mag)
as compared to the 2DG and 1DG fit results as well as $\approx 20 $ mas
systematic differences as a function of position in the field,
when used with a single average point spread function (PSF) for the entire
field of view.
Figure 4 shows an example for the radial difference (2DG-ALS) vs. radius in
frame 57.
Clearly visible slightly elongated images at the edges of many CTIO frames
require field--dependent PSF's to be used in DAOPHOT in order to
obtain better results.
The difficulty is how to relate these PSF's to each other
astrometrically in order to get $x,y$ coordinates in a unique system
for all stars in that frame on the 0.01 pixel level.

In the frame--to--frame comparison again the 2DG shows the best results as
judged from the standard deviations of the transformations of the $x,y$ data.
The 1DG algorithm performs nearly as good as the 2DG in this respect,
while the ALS is clearly inferior.

For comparison with the KPNO results, Figure 2 b) shows a plot of the
fit precision vs. magnitude.
The limiting precision here is only $\approx $ 0.02 pixel.
This can be explained by the poorer image quality of the CTIO
as compared to the KPNO telescope, with variable deviations from the circular
symmetric gaussian image profile depending on the location in the field.
Expressed in arcseconds, both telescopes perform to about the
same level of precision due to the better scale of the CTIO telescope.
Figure 3 b) shows a frame--to--frame comparison, $\sigma_{x}$
vs. magnitude; both frames have been exposed for 600 seconds.
Again a limit of $\approx 5 $ mas in precision is reached for
bright stars in a {\em single} exposure.
The atmosphere seems to be the limiting factor.

\subsection{Basic Mapping Model}

In this chapter the appropriate mapping model between the measured
$x,y$ coordinates of the CCD frames and the corresponding
standard coordinates ($\xi, \eta$) will be investigated.

In order to allow for possible differences in scale and non--orthogonality
of the axis a full linear transformation was adopted as
our basic model. With orthogonal and non--orthogonal terms
separated we have

\[ \xi  = a x + b y + c + e x + f y   \]

\[ \eta = -b x + a y + d + e y - f x  \]

\subsubsection{Optical Distortion Coefficient}

A third--order optical distortion term (D3) was determined from
$x,y$ data of mosaic frames.
A conventional plate adjustment (CPA) of even a field with
many ($\approx 20$) reference stars revealed no significant D3 term.
The mean error on the D3 term is approximately $1.0 \times 10^{-9} \ "/"^{3}$.
In order to obtain a reliable value for the D3 term a procedure
similar to that of the AGK 2 catalog project
(Schorr \& Kohlsch\"utter, \markcite{rmAGK2} 1951)
was followed without the need for reference stars.

Pairs of overlapping CCD frames with offsets in their centers in the
order of half a field size have been transformed onto each other
by extending the linear model with the appropriate D3 term

\[  \Delta \xi  = D3 \ (x_{1} r_{1}^{2} - x_{2} r_{2}^{2} ) \]

\[  \Delta \eta = D3 \ (y_{1} r_{1}^{2} - y_{2} r_{2}^{2} ) \]

with $x_{1},y_{1}$ and $x_{2}, y_{2}$ being the measured coordinates
with respect to the center of distortion on frame 1 and frame 2 respecively and
$ r_{i}^{2} = x_{i}^{2} + y_{i}^{2} \ , \ i=1,2 $.
This algorithm assumes a common distortion term D3 for both frames,
which is very realistic for frames taken shortly after each
other with the same instrument and roughly the same location in the sky.
This assumption was verified by comparing results for D3 obtained
from various frame pairs.

The mean values for D3 and their errors for both telescopes are
given in Table 5 along with the maximal effect per coordinate
on these 2K CCD's.
The D3 term is highly significant and can be determined very
precisely by this method.
For convenience, conversion factors between the different units
for quadratic and third--order terms in the CPA process are
given in Table 6.

\subsubsection{Optical Distortion Center}

A significant offset of the center of distortion (optical axis)
with respect to the geometric center of the CCD frame of $250 \pm 50$ pixel
was found for the CTIO telescope in observing run 4.
A similar offset of $\approx 100 \pm 70$ pixel in another direction
was found for the same telescope in observing run 3, while no such
offsets were found for the KPNO telescope (observing runs 1,2).
Figure 5 a) shows a vector plot of average differences (run 4 - run 3)
of $x,y$ data of field 0743-006 with optical distortion
applied at the geometric frame centers prior to combining the $x,y$ data
of frames for each run.
Figure 5 b) shows the corresponding plot with optical distortion
applied with respect to the optical axis as determined in a pilot
investigation.
In Figure 5 a) there is a systematic error of $\approx 70$ mas
{\em at the frame center} which would have affected all source
positions of that observing run.
According to CTIO staff, such an offset of $\approx 200$ pixel
is within the collimation tolerances of the instrument setup
for each new observing run.

\subsubsection{Tilt Terms}

A difference in tangential points of two overlapping frames
causes tilt terms (p,q) of the form (K\"onig, \markcite{rmK} 1933)

\[  \Delta \xi  = p x^{2} + q xy     \]

\[  \Delta \eta = p xy    + q y^{2}  \]

Similar terms arise when individual frames are not perpendicular
to the optical axis or the location of the tangential point is
uncertain.

A maximum difference in the location of tangential points for
overlapping mosaic frames of 10 arcminutes was used here.
This results in p,q terms as large as $ 1.4 \times 10^{-8} "/"^{2}$
(see Table 6).
A maximum effect of $\Delta x \approx \Delta y \approx 0.01 \ pixel
\ \le \ 7 $ mas  is thus predicted for the edge of the field of view,
which is totally negligible.
Even assuming a tilt of the CCD plane with respect to the focal plane
of the telescope (e.g. due to misalignment of the CCD camera)
of $1^{\circ}$ results in a maximum effect of 0.06 pixel
at the edge of the CCD frame, which would
have little effect on the CPA results.

As expected, no significant p,q terms were found neither in
the $x,y$ transformation of overlapping plates, nor in the CPA of
selected fields.
A CPA with typical secondary reference stars is about a factor of 100
less sensitive to detect p,q terms
as is the $x,y$ data transformation of overlapping frames.

\subsection{External Calibration}

Both instruments show field--dependent systematic errors
of the order of $\approx 20 $ mas  after applying the basic mapping model
including third--order optical distortion (see e.g. Figure 5b).

Unfortunately, presently no external calibration with respect to
a precise reference star catalog can be made.
Plates were taken at the Lick Astrograph with very small
epoch difference (2 months) from some CCD observing runs.
When Hipparcos results will become available, a position
catalog to $\approx 30 $ mas precision for individual stars in
an area of a few square degrees and down to 15th magnitude can
be constructed for these external calibrations of our CCD data.
A rigorous calibration of these FDP's (field distortion patterns) will
then be possible, similar to the procedure used in photographic
astrometry (Zacharias, \markcite{rmFDP} 1995).

\subsection{Optical Positions of Reference Frame Sources}

Here we present position results for 16 sources,
selected as a representative subset of all optical counterparts
on the current list of candidates ($> 400$ sources)
for the extragalactic radio--optical reference frame.
The sources were selected from all 4 CCD
observing runs, sampling all areas in the sky as well as
a wide range in magnitudes.
Problematic cases, e.g. close doubles and sparce fields
were prefered in order to challenge the reduction technique.
Sources with multiple observations (different observing runs)
as well as fields with more than one set of seconday reference stars
were selected in order to obtain accuracy estimates by
external comparisons.

Table 7 gives a summary of the results.
The positions are in the FK5/J2000 system as represented by the IRS catalog,
and based on the 2--dimensional circular gaussian (2DG) fit model.
The radio positions are taken from our radio reference frame
(Johnston \markcite{rmRORF} et al. 1995).
The rms difference (radio--optical) is $\approx 100$ mas for
$\Delta \alpha \cos(\delta)$ and $\Delta \delta$.
Compared to the expected internal errors this only shows
the "wobbles" of the current optical system
and gives no insight into the accuracy of the CCD observations.

A reduction based on a Hipparcos intermediate solution was performed for
the Hipparcos Working Group on Reference Link and shows
an external error consistent with our error estimations.
Detailed results will be published when the final Hipparcos catalog
becomes available.

\subsection{Remarks on Individual Sources}

The following sources are optical doubles, most likely due to a foreground
star. The images of both the extragalactic source and the companion
are consistent with a stellar profile in all cases.
No extended structure, e.g. of an underlying galaxy, was detected.
All these sources are suitable for the radio--optical reference frame link,
at least for now at the 30 mas level.

0153+744 is an optical double with a separation of $\approx 7 \ arcsec$ (10
pixel),
which could be resolved without any problems.

0605-085 is an optical double with a flux of the companion 5 times {\em
brighter}
than the extragalactic source and a separation of 3 arcsec
(4.5 px at KPNO, 7 px at CTIO).
A subtraction of the image profile of the companion was required to obtain
a position of the extragalactic source.

0607-157 is an optical double with a companion 2 times weaker than the
extragalactic source with a separation of 4 arcsec (6 px at KPNO, 10 px at
CTIO).
The cut--out and fit--up procedures (see Section 4) allowed consistent
image profile fits.

0743-006 is an optical double with a companion about a factor of 4 weaker
than the extragalactic source and separated by 2.7 arcsec (6 px at CTIO).
The fit--up and cut--out procedures with the 2DG fit gave
consistent positions within 10 mas, while the {\em allstar} result
was different by $\approx 30$ mas.
Figure 6 shows a contour plot of this source, obtained from a 200--second
exposure CTIO CCD frame in 1.5 arcsec seeing (3.7 pixel FWHM).

1800+440 is an optical double (3 arcsec)
which was successfully fitted with the {\em allstar} algorithm
as well as with the 2DG with use of cut--out and fit--up procedures.
A contour plot of this source has been shown elsewhere
(Zacharias et al. \markcite{rmZ166} 1995).

The large number of optical doubles (5 out of 16 sources) presented here is
not representative for all observations.
These objects were selected for this pilot investigation.

\subsection{Comparison of Multiple Data Sets}

The sources 0336-019, 0605-085, 0607-157 and 0743-006 have been
observed in more than one observing run.
The mean quadratic difference in positions of the same source (see Table 7)
obtained from different runs is $\approx 25 \ mas$,
showing the high accuracy of the CCD observations.
For 0743-006 there are 2 sets of secondary reference stars available
from the Hamburg and Black Birch astrographs respectively.
The agreement between the sets of secondary reference stars is on the
same $\approx 25$ mas level, indicating the high accuracy of the
secondary reference stars and the successful control of
possible magnitude--dependent systematic errors in the astrograph fields.

\section{Discussion}

\subsection{Astrometric Properties of Both Telescopes}

Based on the results of the previous section we estimate some
individual random and systematic error contributions for observations
made with both telescopes.

The error of a single $x,y$ observation depends on the magnitude of the object.
For faint objects the photon statistics limit the
precision and the internal error obtained from the profile fit
or the $x,y$ transformation from frame to frame is a good estimate of
the accuracy of the position.
For bright objects the random errors are as low as
$\approx 0.01$ pixel and the systematic errors dominate.
The astrometrically usable dynamic range depends on the pixel scale
(positions for faint stars are getting worse when undersampled),
as well as on the sky background level.
Although most of our candidates are bright enough to be observed
successfully within the full--Moon period, this has compromised the
usable dynamic range, stressing the
importance of the additional short exposure frames.

Depending on the magnitude of the extragalactic reference link sources,
a typical value
for the random error of an $x,y$ coordinate of a single image on a CCD frame
is $\sigma_{xy} \approx 15 $ mas, but the range was found to be as large as
5 to 31 mas (see Table 7).
Systematic errors as a function of magnitude are expected to be
$\le 10 $ mas  over the magnitude range from secondary reference stars to
extragalactic objects when using the appropriate profile fit model (2DG).
Systematic errors as a function of location on the CCD frame are
found on the $\approx 20 $ mas level, being larger far from the
optical axis but negligible at the frame centers, where the image of the
extragalactic object is usually located.
These systematic errors will average out for different fields because
of the different location of reference stars in each field.
A more rigorous calibration of these FDP's is
in progress which needs to be performed for each observing run separately,
at least for the CTIO telescope.

\subsection{Accuracy Estimate for the Link Procedure}

In this section we will combine all estimates of individual error contributions
for the optical observations of the reference frame link procedure based on
counterparts of extragalactic radio sources.
Special consideration is given to systematic errors in this multi--step
procedure.

We will refer to step 1 as the primary optical reference system.
Options considered here are the IRS and the future Hipparcos and
Tycho catalogs.
By step 2 we denote the secondary reference stars which usually are
obtained photographically by wide--field astrographs.
An intermediate step 3 is sometimes taken with tertiary reference stars
obtained either by Schmidt telesopes or CCD observations with wide fields.
The last step is always the optical observation of the radio source
counterpart itself, either photographically or with CCD.

All error estimates given here are approximate.
The aim is to identify the largest error contribution,
to compare the performance of different options
and to find the limits of this approach to the radio--optical reference
frame link procedure.
All formulae and values to follow are for one coordinate.

As can be seen from the results in Table 7, the standard error of unit
weight $\sigma_{CPA}$ of the adjustment of CCD $x,y$ to secondary reference
star positions can be as low as $\approx 45$ mas,
indicating a sub--micrometer accuracy from a single astrograph plate.
Values for $\sigma_{CPA}$ increase with epoch
difference of the secondary reference star observations and the CCD
observations. This is due to the unknown proper motions in the
secondary reference star data.
No systematic corrections (e.g. galactic rotation) were
applied here.

\subsubsection{Algorithm}

First we will define some quantities to be used in the link of
step $i$ to step $i+1$.

Let $n_{i}$ be the number of stars to be used as reference stars
to link step $i$ to step $i+1$.
The random error of such a link star in step $i$ we denote with $\sigma
ran_{i}$,
and $\sigma sys_{i+1}$ is the systematic error for the link of step $i$ to
$i+1$.
Similarly,
$\sigma xy_{i+1}$ is the precision (random error) of a single $x,y$ observation
for
a link star in step $i+1$,
and $m_{i+1}$ is the number of observations (exposures) for each link star in
step $i+1$.

The random error of a star position obtained in step $i+1$ then is
approximately

\begin{equation}
  \sigma ran_{i+1} \ = \ \frac{\sigma xy_{i+1}}{\sqrt{m_{i+1}}}
\end{equation}

Because there is only a limited number of link stars with associated errors
between the two steps, the link of the system of step $i$ and $i+1$ can not
be made error--free.
The uncertainty in the zeropoint offset, $\sigma z_{i+1}$, between the
coordinate systems
in step $i$ and $i+1$ is approximately

\begin{equation}
  \sigma z_{i+1} \ = \ \sqrt{\frac{\sigma^{2}ran_{i} \ + \ \sigma^{2}ran_{i+1}}
{n_{i}}}
		 \ = \ \frac{\sigma_{CPA}} {\sqrt{n_{i}}}
\end{equation}

Here $\sigma_{CPA}$ is the standard error of unit weight in the least--squares
adjustment
of {\em combined} $x,y$ data of step i+1 to reference star data of step i.
But these formulae hold only for the central area of a frame (plate)
and for a simple (linear) mapping model.
In addition, a factor larger than one is required for other cases,
and a rigorous derivation is given by Eichhorn \& Williams \markcite{rmEW}
(1963).
Our zeropoint offset can be considered as a special case of the
error contribution due to error progapation of the plate constants to field
star
positions.

Finally, the accuracy of a position of an extragalactic link source,
$\sigma_{Q}$ ,
is approximately the rms sum of all zero point offsets from previous steps
plus the systematic errors and the precision, $\sigma ran_{Q}$,
of all optical observations of the source itself

\begin{equation}
 \sigma_{Q} \ = \ \sqrt{\sigma^{2}ran_{Q} \ + \
		  \sum_{i=2}^{k} \sigma^{2}z_{i} \ + \
		  \sum_{i=2}^{k} \sigma^{2}sys_{i}}
\end{equation}

With systematic errors we mean here errors not averaging out with the number
of stars used for the link of a single extragalactic source.
It is assumed that such systematic errors (e.g. depending on magnitude for a
particular plate) will be different for different extragalactic source fields
and thus (at least partly) random when results for many sources are combined.
Systematic errors inherent in this technique and not averaging out with
different fields can't be investigated here.
An external comparison with other methods for the extragalactic link
procedure will be made in the future.

\subsubsection{Accuracy Estimate for the Secondary Reference Star Positions}

Here we will start out with 3 options for the primary optical reference system
(step 1) and discuss 12 cases for determining secondary reference star
positions
(step 2).
All cases are summarized in Table 8.

The currently available IRS system has a density of $\approx 0.9 \
stars/degree^{2}$
and a precision of $\approx 200 $ mas for epochs of 1980 to 1994, where most
of our data were taken.
The usable field of views for the Hamburg Zone Astrograph (ZA),
the Black Birch Astrometric Observatory (BBAO) astrograph and the Lick
Astrograph
are approximately 36, 25 and 9 $degree^{2}$ respectively.
For the Hipparcos catalog we assume a mean $\sigma ran_{1} = 10 $ mas
for the epoch range of our data.
After the Tycho catalog is combined with the Astrographic Catalog (AC)
data in order to obtain proper motions, we assume
a mean $\sigma ran_{1} = 50 $ mas for that catalog at the epoch of our data.

Systematic errors depending on magnitude are controlled with a diffraction
grating at the astrographs.
Preliminary results indicate magnitude terms on the
order of $0 \ to \ 1 \ \mu m $ per 5 magnitudes, this is up to $20 $ mas/mag.
The error on determining this term is about $ 2 $ mas/mag.
This is a systematic error for a plate or field, which varies from
field to field.

All cases 4 are based on a $ 1 \ deg^{2}$ CCD frame.
Cases 4b and 4c assume a mini--block adjustment of an area of $\approx 9 \
deg^{2}$.
As can be seen from Table 8, a considerable improvement will be gained
from the Hipparcos catalog as compared to the current IRS.
A further improvement can be obtained by using the Tycho catalog,
but this means a tremendous effort on plate measuring and the availability
of the AC in order to derive good proper motions to be combined with
the original Tycho observations.
This is only worth the effort if the systematic errors can be controlled
to this level.

The CCD option is only competitive here when used with block adjustment
techniques, at least in local fields.
Because of the expected lower systematic errors, e.g. as a function of
magnitude, this may be the way to go in the future.
Such CCD observations could be based directly on Hipparcos stars with
about the same precision as could be obtained from a Tycho--based
solution, thus excluding possible systematic errors from the
Tycho proper motions.

\subsubsection{Position Accuracy of the Extragalactic Sources}

With the algorithm as above and a typical internal precision of
a CCD observation of $\sigma xy \approx 15 $ mas and $m = 2$
observations per source, we have a random error for the optical
position of an extragalactic source of $\sigma ran_{Q}  \approx 11 $ mas.

Random errors from the CCD observations of the link stars
(secondary reference stars) are even smaller due to their brightness,
a typical value is $\sigma_{fit} \le 10 $ mas.
The largest error contribution here is the influence due to the
atmosphere in case of the short exposure frames.
According to Lindegren \markcite{rmLin} (1980) this amounts to
$\sigma xy_{3} \approx 30 $ mas for a single exposure;
with $m_{3} \approx 4$ we have according to Eq. (1)
$\sigma ran_{3} \approx 15 $ mas.

Let us assume a random error from the astrograph observations of
a link star (secondary reference star) of
$\sigma ran_{2} \approx 70 $ mas.
With $n_{2} \approx 15$ stars for that link
between step 2 (secondary reference stars) and
step 3 (CCD observations) we thus obtain according to Eq. (2)
a zeropoint error of $\sigma z_{3} \approx 20 $ mas.
This strongly depends on the number of stars used and the epoch
difference between the CCD and astrograph observations.
Individual results of the precision and accuracy properties of the
CCD observations can be found in Table 7.

Putting everything together, and assuming
$\sigma sys_{2} \approx \sigma sys_{3} \approx 10 $ mas,
we expect an accuracy of the optical
position of an extragalactic object to be in the order of
$\sigma_{Q} \approx 30 $ mas plus $\sigma z_{2}$ as discussed
in the previous section, which is negligible in case of
Hipparcos--catalog--based secondary reference stars
($\sigma z_{2} \approx 5 $ mas).
When using the IRS catalog we have $\sigma z_{2} \approx 45 $ mas
plus $\sigma sys_{1} \approx 100 $ mas.
Individual estimates of $\sigma_{Q}$ (based on the Hipparcos catalog)
are given in Table 7 for each object.
Thus currently the largest error contribution comes from the primary system,
the FK5, as represented by the IRS.
With the use of the Hipparcos results the largest error contribution
comes from the weak link of the secondary reference stars to the
CCD observations due to the small field of view of the CCD's and
the relatively poor limiting magnitude of the secondary reference stars.
A large epoch difference between the secondary reference star
and the QSO observations significantly increases the noise in
this crucial step, regardless of any additional possible systematic
errors indroduced by unknown proper motions.

\subsection{Comparison to Other Investigations}

Other major procedures for the position link of the radio and optical
reference frame are the HST (Hubble Space Telescope)
observations of selected pairs of Hipparcos stars and bright extragalactic
candidates and the VLBI/VLA observations of Hipparcos radio stars.
The HST observations are of higher internal precision than our observations
but are not so numerous ($\approx 40$ pairs) and depend on the absolute
calibration of the FGS fields.
The radio star approch is very precise and direct but is based only on
less than 10 objects which are not well distributed over the sky.

Our approach contributes significantly to the link process and allows the
important check to be made on possible systematic errors of the other methods.

The Sloan Digital Sky Survey (SDSS) will be helpful in order to
densify the grid of secondary (and tertiary) reference stars
in the galactic north pole region.
The positional accuracy of the optical counterparts of the
extragalactic radio sources from SDSS will be inferior to our observations
due to shorter integration time.

\section{Conclusions}

The feasibility of this approach to the radio--optical reference frame link
has been proven here using wide field CCD observations
with the KPNO and CTIO 0.9 m telescopes.

The link to the primary reference star system, as represented by the
Hipparcos catalog in the near future,
is based entirely on photographic plates obtained with dedicated
astrographs in both hemispheres.
A deeper limiting magnitude and higher precision for the link stars
is most important now, and CCD observations at the astrographs are in progress
to provide more reference star positions.
These observations will also provide an additional determination
of possible magnitude--dependent systematic errors in the entire
procedure.

The 0.9 m telescope CCD observations have acquired a huge amount of
high precision optical observations of extragalactic sources within
a short period of time.
The precision for a single long exposure
is in the range of 5 to 31 mas (average $\approx 15$ mas)
depending on the magnitude of the object.
Field--dependent systematic errors exist on a 20 mas level,
but they will be externally calibrated in the near future.
A 20 mas precision level was reached previously with prime focus
photography, but only for few objects per year at large telescopes.

With 3 more CCD runs we hope to complete observations at the 0.9 m
telescopes for about 400 sources which would allow a position tie to
the Hipparcos system on the 1 mas level.

In addition to the position information, a {\em structure} analysis
of the optical counterparts is highly desirable.
Because these objects have already been selected to be
compact, the search for structure has to be made
with much higher resolution than the 0.9 m ground--based
telescopes can offer.
Optical interferometry, adaptive optics or the HST are the only options
at the moment.
Currently our structure analysis is limited to identify suitable
candidates for the link process, i.e. optical sources
free of nearby disturbing foreground stars and galaxies.

After a sufficiently rigid link between the radio and optical
systems has been established, our observations will be used
to identify outliers which will have astrophysical implications
about the nature of these compact extragalactic objects.
The biggest advantage of our approach to the extragalactic
reference frame link is the large number of sources involved.
Observations can easily be maintained in the future from
ground--based telescopes, providing also an epoch
difference large enough for a proper motion tie of the
Hipparcos system within the next decade.
With only minor improvements and more observations
a much higher precision and accuracy can be reached by
this technique in the near future.

Position results of a large number of sources will be
published after the Hipparcos catalog becomes available.
No conclusions should be drawn from the positions published
here based on the IRS system.
When using a Hipparcos intermediate solution, as required
for the Hipparcos Working Group on Reference Link,
positional results with respect to the radio frame
are in agreement with the error estimation given above.

Ultimately a space mission like
FAME (Johnston \markcite{rmFAME} 1995)
or GAIA (Perryman \& van Leeuwen \markcite{rmGAIA} 1996)
will provide optical positions for some (in case of FAME)
or most (in case of GAIA) of these objects with an accuracy
better than current VLBI radio observations.
Until then we will hopefully have a much better understanding
of the astrophysical and astrometric properties of
these objects in order to be able to concentrate on the
most suitable candidates for a reference frame.

\acknowledgments

Chr. de Vegt wishes to thank the
Bundesministerium f\"ur Forschung und Technologie (BMFT)
for financial support under Grant No. 50008810 (Hipparcos).
We further thank J.L. Russell and M.I. Zacharias for assistance
with observing, as well as J. M\"unkel for
assistance with the astrographic plate measuring and reduction
process.

\newpage

\begin{figure}
\caption{Difference in x--coordinate for 2--dimensional minus
   1--dimensional gaussian fit from the same pixel data
   vs. instrumental magnitude, example for
   \ a) KPNO, \ b) CTIO.
   1 dot represents the mean of 4 differences.}
\end{figure}

\begin{figure}
\caption{Fit error for x--coordinate of individual stars
     vs. instrumental magnitude; example for
     \ a) KPNO, \ b) CTIO.}
\end{figure}

\begin{figure}
\caption{Standard error of unit weight for x--coordinate of a
	 frame--to--frame transformation with the same tangential point
	 vs. instrumental magnitude; example for
         \ a) KPNO, \ b) CTIO.
   1 dot represents the mean of 4 differences.}
\end{figure}

\begin{figure}
\caption{Radial difference for 2--dimensional gaussian fit
	 minus point spread function fit from the same pixel data
	 vs. radial distance from the CCD center; an example from CTIO data.}
\end{figure}

\begin{figure}
\caption{Vector plot for $x,y$ position differences of the field
	 0743-006 as observed in Dec.94 minus as observed in Febr.95
	 with the same CTIO telescope, with optical distortion corrected at
	 \ a) the geometrical frame center,
	 \ b) the assumed location of the optical axis.
	 The differences are increased by a scale factor of 2000.}
\end{figure}

\begin{figure}
\caption{Contour plot of 0743-006 from a 200 seconds CCD
         exposure taken at CTIO. The image of the QSO is the brighter of the
two objects.}
\end{figure}

\newpage

\begin{table}
\caption{Properties of instruments (telecopes and CCD's). }
\begin{tabular}{lccl}
		&   KPNO         &   CTIO   &        \\
\tableline
aperture        &   0.9          &   0.9    &  meter \\
focal ratio     &  f/7.5         &  f/13.5  &        \\
optical system  & R/C + corr.    & cassegrain &      \\
plate scale     &  28.3          &  17.0    &  "/mm  \\
pixel size      &  24.0          &  24.0    &  $\mu m$ \\
pixel scale     &   0.68         &   0.40   &  "/pixel  \\
field of view   &  23.1          &  12.8    &  '     \\
readout time    &  140           &  70      & seconds\\
\end{tabular}
\end{table}

\begin{table}
\caption{Summary of observing runs.}
\begin{tabular}{lcccc}
              run number   &     1         &     2        &     3       &  4 \\
\tableline
date                       &  April 94     &  Oct. 94     &  Dec. 94    & Feb.
95 \\
observatory                &   KPNO        &  KPNO        &  CTIO       & CTIO
\\
number of usable nights    &   5           &   5          &   7.5       &   8
\\
average seeing (arcsec)    &  1.6 ... 3    &  1.4 ... 1.8 & 1.2 ... 1.8 & 1.2
... 2.0 \\
number of observed sources &   35          &   70         &   60        &   68
\\
number of object frames    &  103          &  163         &  117        &  150
\\
number of short exp. frames&   89          &  160         &  238        &  340
\\
number of test frames      &   61          &    6         &   23        &   13
\\
\end{tabular}
\end{table}

\newpage

\begin{table}
\caption{Results from frame \& fit model to frame \& fit model comparisons of
KPNO test field 1656+053.
	 Standard errors $\sigma_{x}, \sigma_{y}$ are given in milli--pixel.}
\begin{tabular}{clclrrl}
frame 1  &  fit model 1 & frame 2 & fit model 2 &  $\sigma_{x} $  &
$\sigma_{y} $  &  remark  \\
\tableline
 68 & 2DG & 68  &  ALS  &  22  &  23  &  long exposure \\
 68 & 2DG & 68  &  1DG  &  28  &  24  &  \\

 74 & 2DG & 74  &  ALS  &  18  &  20  &  long exposure \\
 74 & 2DG & 74  &  1DG  &  19  &  20  &  \\

 71 & 2DG & 71  &  1DG  &  29  &  30  &  magnitude equation \\
 71 & 2DG & 71  &  ALS  &  39  &  36  &  ok \\

 72 & 2DG & 72  &  1DG  &  29  &  24  &  magnitude equation \\
 72 & 2DG & 72  &  ALS  &  32  &  31  &  ok \\

 73 & 2DG & 73  &  1DG  &  27  &  34  &  magnitude equation \\
 73 & 2DG & 73  &  ALS  &  20  &  22  &  ok\\
\tableline
 68 & 2DG & 74  &  2DG  &  35  &  35  & ok long - long \\
 68 & 1DG & 74  &  1DG  &  45  &  41  &      \\
 68 & ALS & 74  &  ALS  &  44  &  41  &      \\

 74 & 2DG & 73  &  2DG  &  47  &  55  & ok long - short \\
 74 & 1DG & 73  &  1DG  &  72  &  72  &  \\
 74 & ALS & 73  &  ALS  &  63  &  75  &  \\

 73 & 2DG & 72  &  2DG  &  62  &  44  &  \\
 73 & 1DG & 72  &  1DG  &  70  &  55  &  \\
 73 & ALS & 72  &  ALS  &  78  &  61  &  \\

 73 & 2DG & 71  &  2DG  &  64  &  73  &  \\
 73 & 1DG & 71  &  1DG  &  73  &  77  &  \\
 73 & ALS & 71  &  ALS  &  84  &  81  &  \\
\end{tabular}
\end{table}

\newpage

\begin{table}
\caption{Results similar to Table 3 for the CTIO test field $0646-306$.}
\begin{tabular}{clclrrl}
frame 1  &  fit model 1 & frame 2 & fit model 2 &  $\sigma_{x} $  &
$\sigma_{y} $  &  remark  \\
\tableline
 54 & 2DG & 54 & 1DG & 30 & 36 & long exposure\\
 54 & 2DG & 54 & ALS & 50 & 63 & \\

 57 & 2DG & 57 & 1DG & 38 & 40 & short exposure\\
 57 & 2DG & 57 & ALS & 39 & 58 & \\

 58 & 2DG & 58 & 1DG & 31 & 37 & short exposure\\
 58 & 2DG & 58 & ALS & 48 & 75 & \\
\tableline
 53 & 2DG & 54 & 2DG & 46 & 43 & long exposure\\
 53 & 1DG & 54 & 1DG & 49 & 48 & \\
 53 & ALS & 54 & ALS & 75 & 89 & fit model not ok\\

 54 & 2DG & 57 & 2DG & 62 & 60 & \\
 54 & 1DG & 57 & 1DG & 80 & 73 & \\
 54 & ALS & 57 & ALS & 91 & 96 & \\

 57 & 2DG & 58 & 2DG & 68 & 80 & \\
 57 & 1DG & 58 & 1DG & 80 & 83 & \\
 57 & ALS & 58 & ALS & 101&128 & \\

 57 & 2DG & 60 & 2DG & 74 & 79 & \\
 57 & 1DG & 60 & 1DG & 82 & 90 & \\
 57 & ALS & 60 & ALS &106 &107 & \\

\end{tabular}
\end{table}

\newpage

\begin{table}
\caption{Summary on third--order optical distortion of both telescopes}
\begin{tabular}{lrrl}
		              &   KPNO   &   CTIO     &  unit \\
\tableline
mean value of D3              &    50    &   134      &  $"/"^{3}$ \\
mean value of D3              &$-0.49 \times 10^{-9}$ & $ -0.46 \times 10^{-9}$
& $px / px^{3}$ \\
standard error on D3          &$ 0.02 \times 10^{-9}$ & $  0.03 \times 10^{-9}$
& $px / px^{3}$ \\
maximum effect at frame edge  &  0.53    &   0.49   &  px  \\
scale                         &  0.68    &   0.40   &  "/px \\
\end{tabular}
\end{table}

\begin{table}
\caption{Conversion factors for quadratic and third--order terms,
          {\em scale} in arcsecond (") per pixel (px) and
	  C1 =  3600 * 180 / $\pi \ \approx \ $ 206264.8 }

\begin{tabular}{llll}
  term             &   input units &  output units  &   factor  \\
\tableline
 quadratic (p,q)   & $px / px^{2}$ & $ " / "^{2}$   &  1/scale  \\
 quadratic (p,q)   & $" / "^{2}$   & $ rad/rad^{2}$ &  C1   \\
 tilt angle(p,q)   & $rad/rad^{2}$ &   "            &  C1   \\
\tableline
 distortion (D3)   & $px / px^{3}$ & $ " / "^{3}$   & $1/scale^{2}$\\
 distortion (D3)   & $" / "^{3}$   & $ rad/rad^{3}$ & $ C1^{2}$ \\
 distortion (D3)   & $" / "^{3}$   & $ " / deg^{3}$ & $3600^{3}$  \\
\end{tabular}
\end{table}

\newpage

\addtocounter{table}{1}

\begin{table}
\caption{Accuracy of secondary reference stars.
	 Additional errors due to unknown proper motions of the
	 anonymous secondary reference stars are not
	 included here.
	 The Tycho catalog is assumed to have been complemented
	 with ground--based observations (AC) in order to contain
	 reliable proper motions.
	 Case 4b and 4c assume a block adjustment
	 of overlapping frames in a $9 \ deg^{2}$ area (mosaic CCD).
	 See Section 6.2.1 for explanation of columns.}

\begin{tabular}{lllrrrrrrr}
case  & primary & $\sigma ran_{1}$ & $\sigma sys_{1}$ & instr. & $n_{1}$ &
        $\sigma xy_{2}$  & $m_{2}$ & $\sigma ran_{2}$ & $\sigma z_{2}$ \\
      & catalog &   mas            &   mas            &        &         &
	   mas           &         &   mas         &    mas         \\
\tableline
 1 a  & IRS &  200 & 100 & ZA   &  32 & 90 &  4 &  45 & 36.0  \\
 1 b  &     &      &     & BBAO &  22 &100 &  4 &  50 & 44.0  \\
 1 c  &     &      &     & Lick &   8 & 50 &  3 &  29 & 71.0  \\
\tableline
 2 a  & Hip &   10 &   1 & ZA   & 100 & 90 &  4 &  45 &  4.6  \\
 2 b  &     &      &     & BBAO &  70 &100 &  4 &  50 &  6.1  \\
 2 c  &     &      &     & Lick &  25 & 50 &  3 &  29 &  6.1  \\
\tableline
 3 a  &Tycho&   50 &   5 & ZA   &1000 & 90 &  4 &  45 &  2.1  \\
 3 b  &     &      &     & BBAO & 700 &100 &  4 &  50 &  2.7  \\
 3 c  &     &      &     & Lick & 250 & 50 &  3 &  29 &  3.7  \\
\tableline
 4 a  &Tycho&   50 &   5 &  CCD &  25 & 30 &  4 &  15 & 10.4  \\
 4 b  &     &   50 &   5 & mCCD & 225 & 30 &  4 &  15 &  3.3  \\
 4 c  & Hip &   10 &   1 & mCCD &  22 & 20 &  4 &  10 &  3.0  \\
\end{tabular}
\end{table}

\end{document}



\begin{deluxetable}{lclccccrlcrcrclrclrrccc}
\small
\tablecolumns{23}
\tablewidth{0pc}
\tablenum{7}
\tablecaption{Position Results of Selected Sources}
\tablecomments{
	 t = type of object with Q = QSO, L = BL Lac;
	 mag = approx. magnitude;
	 run = observing run number (see Table 2);
	 nCCD = number of good CCD frames, l = long, s = short exposure;
	 FW = mean full width at half maximum of image profiles in pixel;
	 $\sigma_{xy}$ = fit error from a single long exposure in milliarcseconds
(mas);
	 secondaries = information about secondary reference stars;
	 tel = telescope (ZA for Hamburg Zone Astrograph, BY for Yellow lens BBAO);
	 yr = mean epoch in years--1900, nref = number of reference stars used,
	 with letter $m$ indicating mosaic frames;
	 $\sigma_{CPA}$ = standard error of unit weight of CPA in mas;
	 RA,DC = optical J2000 position in the IRS/FK5 system;
         opt.$-$radio = $\Delta \alpha \cos \delta$,  $\Delta \delta$ with
respect
	     to the radio position of Johnston et al.(1995);
         $\sigma_{Q}$ = estimate of the accuracy of the optical position if
based on
		 the upcoming Hipparcos catalog and assuming
		 $\sigma sys_{2} = \sigma sys_{3} = 10 $ mas;
         sep = separation in arcseconds if optically double;
         flr = flux ratio object/companion.}

\tablehead{
   \colhead{source}   & \colhead{t}  & \colhead{mag}  & \colhead{run}
 & \multicolumn{2}{c}{nCCD}          & \colhead{FW}   & \colhead{$\sigma_{xy}$}
 & \multicolumn{3}{c}{secondaries}   & \colhead{$\sigma_{CPA}$}
 & \multicolumn{3}{c}{RA (2000)}     & \multicolumn{3}{c}{DC (2000)}
 & \multicolumn{2}{c}{opt.$-$radio}  & \colhead{$\sigma_{Q}$}
 & \colhead{sep}      & \colhead{flr}  \\
   \multicolumn{4}{c}{ }             & \colhead{l}    & \colhead{s}
 & \colhead{px}       & \colhead{mas}& \colhead{tel}  & \colhead{yr}
 & \colhead{$n_{ref}$}& \colhead{mas}& \colhead{h}
 & \colhead{m}        & \colhead{s}  & \colhead{$^\circ$}
 & \colhead{'}        & \colhead{"}  & \multicolumn{2}{c}{mas}
 & \colhead{mas}      & \colhead{"}  & \colhead{} }
\startdata

 0153$+$744&Q&16   &2 &2 &4 &2.5 &11 &ZA&79.8 &31\phantom{ m}&    147
&01&57&34.9189&$+$74&42&43.221&$-$183 &$-$10 & 31 & 6.9 &0.8 \nl \nl

 0308$-$611&Q&18   &3 &2 &6 &3.4 & 9 &BY&90.4 &10 m          &\phn 82
&03&09&56.1035&$-$60&58&39.063&  31 & $-$7 & 31 &     &    \nl \nl

 0336$-$019&Q&18.4 &2 &3 &2 &2.3 &15 &BY&89.9 &13\phantom{ m}&\phn 90
&03&39&30.9368&$-$01&46&35.772& $-$15 & 31 & 30 &     &    \nl
 0336$-$019& &     &3 &3 &0 &3.8 & 8 &BY&89.9 & 6\phantom{ m}&\phn 62
&03&39&30.9341&$-$01&46&35.772& $-$55 & 31 & 30 &     &    \nl \nl

 0605$-$085&Q&18.5 &2 &3 &2 &2.3 &17 &BY&91.2 &23\phantom{ m}&\phn 63
&06&07&59.6832&$-$08&34&50.138&$-$237&$-$161 & 22 & 3.1 &0.2 \nl
 0605$-$085& &     &3 &4 &5 &4.0 &16 &BY&91.2 &13 m          &\phn 69
&06&07&59.6809&$-$08&34&50.111&$-$271&$-$134 & 26 &     &    \nl \nl

 0607$-$157&Q&18   &2 &2 &2 &2.4 & 9 &BY&89.9 &27\phantom{ m}&\phn 53
&06&09&40.9356&$-$15&42&40.672&$-$201&   5 & 19 & 4.3 &1.9 \nl
 0607$-$157& &     &4 &3 &5 &4.0 &12 &BY&89.9 &17 m          &\phn 52
&06&09&40.9337&$-$15&42&40.690&$-$228& $-$13 & 21 &     &    \nl \nl

 0629$-$418&Q&19.3 &3 &2 &5 &3.2 &10 &BY&90.1 &23 m          &\phn 99
&06&31&11.9898&$-$41&54&27.071& $-$92&$-$126 & 26 &     &    \nl \nl

 0646$-$306&\nodata&\nodata&3 &2 &6 &3.5 &22 &BY&90.0 &37 m  &\phn 67
&06&48&14.0888&$-$30&44&19.641& $-$98 & $-$5 & 24 &     &    \nl \nl

 0743$-$006&L&17.1 &3 &2 &5 &3.7 & 9 &BY&91.1 &30 m          &\phn 78
&07&45&54.0774&$-$00&44&17.746 &$-$73&$-$208 & 22 & 2.7 &3.8 \nl
 0743$-$006& &     &  &2 &5 &    & 9 &ZA&91.0 &24 m          &\phn 71
&07&45&54.0798&$-$00&44&17.742 &$-$38&$-$204 & 22 &     &    \nl
 0743$-$006& &     &4 &3 &5 &4.0 & 9 &BY&91.1 &47 m          &\phn 95
&07&45&54.0780&$-$00&44&17.718 &$-$64&$-$180 & 21 &     &    \nl
 0743$-$006& &     &  &3 &5 &    & 9 &ZA&91.0 &35 m          &\phn 59
&07&45&54.0791&$-$00&44&17.714 &$-$48&$-$176 & 19 &     &    \nl \nl

 0818$-$128&L&15   &3 &2 &6 &3.3 & 5 &BY&90.2 &20 m          &\phn 64
&08&20&57.4467&$-$12&58&59.173& $-$14 & $-$6 & 21 &     &    \nl \nl

 0906$+$015&Q&17.8 &1 &2 &1 &2.7 &20 &BY&91.3 &12\phantom{ m}&\phn 64
&09&09&10.0844&$+$01&21&35.379&$-$108&$-$240 & 28 &     &    \nl
 0906$+$015& &     &4 &2 &4 &3.6 &13 &BY&91.3 & 7 m          &\phn 45
&09&09&10.0853&$+$01&21&35.434& $-$94&$-$185 & 24 &     &    \nl \nl

 1004$+$141&Q&19   &1 &6 &0 &3.0 &31 &ZA&87.7 &19\phantom{ m}&    115
&10&07&41.4922&$+$13&56&29.482& $-$85&$-$120 & 33 &     &    \nl \nl

 1039$+$811&Q&16.5 &1 &2 &2 &3.0 &24 &ZA&84.2 &17\phantom{ m}&    162
&10&44&22.9495&$+$80&54&39.557&$-$268& 114 & 45 &     &    \nl \nl

 1111$+$149&Q&18   &1 &2 &1 &2.9 &22 &ZA&90.3 & 8\phantom{ m}&\phn 84
&11&13&58.6968&$+$14&42&26.769&  25&$-$184 & 37 &     &    \nl \nl

 1656$+$053&Q&16.5 &1 &2&18 &3.0 & 9 &ZA&86.4 &23 m          &\phn 91
&16&58&33.4388&$+$05&15&16.370&$-$123& $-$75&  25 &     &    \nl \nl

 1800$+$440&Q&16.8 &1 &3& 1 &3.2 &11 &ZA&91.5 &25\phantom{ m}&\phn 70
&18&01&32.3140&$+$44&04&22.049&  $-$8& 147&  21 & 2.9 &1.6 \nl \nl

 2201$+$315&Q&15.6 &2 &2& 2 &2.4 &13 &ZA&87.8 &48\phantom{ m}&\phn 75
&22&03&14.9825&$+$31&45&38.293&  85&  23&  21 &     &
\enddata

\end{deluxetable}